\documentclass[useAMS,usenatbib]{mn2e}
\usepackage{graphicx}
\usepackage[latin1]{inputenc}
\usepackage{color}
\usepackage{times}
\usepackage{natbib}
\usepackage{setspace}
\usepackage{amsmath}
\usepackage{amsfonts}
\usepackage{amssymb}
\usepackage{multirow}
\usepackage{subfigure}
\usepackage{epstopdf}

\newif\ifAMStwofonts
\AMStwofontstrue
\definecolor{red}{rgb}{1,0.,0.}

\title[The impact of global environment on galaxy mass functions] {The impact of global environment on galaxy mass functions at low redshift}
\author[Calvi et al.]{Rosa Calvi$^{1}$\thanks{E-mail:
email: rosa.calvi@unipd.it}, Bianca M. Poggianti$^{2}$, Benedetta Vulcani$^{1,2,3}$, Giovanni Fasano$^{2}$\\
  $^1$Astronomical Department, Padova University, Italy \\
  $^2$INAF-Astronomical Observatory of Padova, Italy\\
  $^3$Kavli Institute for the Physics and Mathematics of the Universe, University of Tokyo, Kashiwa, 277-8582, Japan\\
}

\begin{document}

\date{Accepted ... Received ...}

\pagerange{\pageref{firstpage}--\pageref{lastpage}} \pubyear{}

\maketitle
\label{firstpage}

\begin{abstract}
We study the galaxy stellar mass function in different environments in
the local Universe, considering both the total mass function and that
of individual galaxy morphological types. We compare the mass
functions of galaxies with $\rm log_{10} M_{\star}/M_{\odot} \geq
10.25$ in the general field and in galaxy groups, binary and single
galaxy systems from the Padova-Millennium Galaxy and Group Catalogue
at $z=0.04-0.1$ with the mass function of 
galaxy clusters of the WIde-field Nearby
Galaxy-Cluster Survey at $z=0.04-0.07$.  Strikingly, the variations of
the mass function with global environment, overall, are small and
subtle. The shapes of the mass functions of the general field and
clusters are indistinguishable, and only small, statistically
insignificant variations are allowed in groups. Only the mass function
of our single galaxies, representing the least massive haloes and
comprising less than a third of the general field population, is
proportionally richer in low-mass galaxies than other environments.
The most notable environmental effect is a progressive change in the
upper galaxy mass, with very massive galaxies found only in the most
massive environments. This environment-dependent mass cut-off is
unable to affect the Schechter parameters and the K-S test, and can
only be revealed by an ad-hoc analysis.
Finally, we show how, in each given environment, the mass function changes with
morphological type, and that galaxies of the same morphological type
can have different mass functions in different environments.
\end{abstract}


\section{Introduction}
\label{s:introduction}

The formation and evolution of galaxies is a subject of
great complexity.  If on the one hand the cosmological framework of
the structure formation in the Universe is well in place and in
agreement with observations, on the other hand the significant
differences that emerge between galaxy models and the observed trends
leave open many unsolved problems and intriguing scenarios on the
physical processes involved.

Studies of the galaxy mass function (MF hereafter) are expected to
provide important clues in this context because the MF, in addition to
constraining the baryonic content of galaxies, is the result of the
hierarchical mass assembly of dark matter halos, the intrinsic
physical processes and the transformation processes that galaxies
experience during their lifetimes \citep{BR}.

In recent years, the combination of spectroscopic catalogs with large
photometric (especially infrared) surveys such as 2dF, SDSS, 2MASS has
made possible to explore the distribution of galaxy stellar masses in
the local universe over the mass range $\rm M_{\star} = 10^{7}-10^{12}
M_{\odot}$ \citep{Yang,Li,BA4} and as a function of the galaxy
environment. There
are two main ways to characterize the environment. One is to refer to the
``global'' structure to which a galaxy belongs (from clusters, to
groups, to lower mass haloes). The second one
 is based on ``local'' galaxy density which can be
parametrized following different techniques, for example by the number
density of objects within some distance or measuring the distance of a
galaxy to the $\rm N_{th}$ nearest neighbour, with $\rm N_{th}$
typically between 5-10.

Various studies have found that the galaxy MF
depends on local galaxy density, with denser environments hosting on
average more massive galaxies, both in the nearby universe \citep{Bam,BA2,VU3}
and at higher redshifts \citep{BOL,VU3}.
These works have analyzed galaxies in the general field, measuring
the galaxy local density based on the 5th nearest neighbours.
\citet{VU3} have also shown the importance of the local density
in regulating the mass distribution
within galaxy clusters at low and intermediate redshift.

In contrast, the variation of the MF with global environment, and
consequently halo mass, has not been fully understood.
In the local universe, the study from \citet{Bal01} finds a
different mass function in clusters compared to the field, while other
works detect no difference \citep{vdL,MER10}.

The MF in distant clusters and its
evolution from $\rm z\sim 0.8$ to $\rm z\sim 0$ have been analyzed
for the first time in \citet{VU1}.
They found that the MF evolves with redshift: clusters at high-z
show proportionally more massive galaxies than clusters at low-z,
probably as a consequence of the mass growth of
galaxies due to star formation in both cluster galaxies and, most of
all, in galaxies infalling from the cluster surrounding
areas.

A preliminary comparison with the field MFs
taken from literature did not find evidence for an environmental mass
segregation.

Performing an additional study on cluster and field data at $\rm
0.3\leq z\leq 0.8$, \citep{VU2} found that the mass distribution at
these redshifts does not show a dependence on global environment,
being the global environment defined as clusters, groups or general
field. As a consequence, the evolution of the MF between $z \sim 0.8$
and today is similar in clusters and the general field.
Differences in the MF of galaxies at intermediate
redshifts become evident when comparing group and isolated
galaxies, as found for the the zCOSMOS sample by \citet{KOV10}.
At similar redshifts, \citet{GIO} analysing a sample of X-ray galaxy
groups, found that the MF of passive
galaxies shows a difference from groups to field while the
star-forming MF is similar in all environments.

The dependence of the MF on local density and its invariance from
clusters to groups to general field \citep{VU3,VU2} led these authors
to conclude that at least at $\rm z\leq 0.8$ local density is more
important than global environment in determining the galaxy stellar
mass distribution, suggesting that galaxy properties are not much
dependent of halo mass, but do depend on local scale processes.

On the theoretical side, \citet{Mo} found a correlation
between the stellar mass of the central galaxy and the mass of the
dark matter halo in N-body simulations. Indeed, observationally, at
cluster scales the mass distribution of central galaxies appears to be
a function of halo mass \citep{Yang} but whether the total mass
function depends on global environment is still an open question
for both observations and simulations.
As far as the dark matter component is concerned, high resolution
numerical simulations predict
that dark matter haloes contain a population of subhaloes whose mass
function is found to be universal, independently of the mass of the host
halo \citep{GIOC}. Several authors have found that simulations
coupled with semi-analytic models are not able to reproduce the
mass function of low-mass galaxies (e.g. \citet{Font,GUO}). 
A detailed theoretical investigation for
different halo masses has not been carried out yet and is currently
underway (Vulcani et al. in prep.).

In this work we analyze the MF of galaxies at low redshift
($\rm 0.04-0.1$) as a function of ``global'' environment. We study both the
general field MF and, for the first time, its variation in
progressively less massive ``haloes'' from clusters, to groups, to
binary systems and single galaxies, covering a range of system masses
from $10^{15} M_{\odot}$ to systems that are expected to be of the
order of a few times $10^{12} M_{\odot}$.  Our aim is to understand if
and how the MF varies with the global environment in the local
Universe, where we are able to perform a detailed analysis isolating
also low mass environments.


In addition, we study the MF of different morphological
types: ellipticals, lenticulars and later-type galaxies.
Our aim here is twofold: to characterize the differences in MF
between a morphological type and the other, in each given environment,
and to investigate whether the MF of a given type changes with
environment.


The paper is structured as follows: after presenting our datasets in \S2,
we present our mass measurements and methods in \S3.1 and 3.2, respectively.
\S3.3 shows a comparison of our general field MF with
previous literature results.
Our main results are presented in \S4.
The galaxy MFs by morphological types, how they differ
with environment and from each other, are given in \S5.
We discuss our results in \S6 and summarize them in \S7.
Throughout this paper we consider a
$\rm \Lambda CDM$ cosmology with
$\rm \Omega _{M}=0.3$, $\rm \Omega _{\Lambda}=0.7$ and Hubble constant
of $\rm H_{0} = 70 \, km \, s^{-1} \, Mpc^{-1}$, a Kroupa (2001) IMF
and Vega magnitudes.

\section{Galaxy samples at low-z}

In order to present a complete overview of how galaxies properties
vary in different environments, we used two galaxy samples in the
local universe: group, binary, single and, all together, general field
galaxies were selected from the Padova Millennium Galaxy and Group Catalog
(PM2GC) \citep{Ca}, while cluster galaxies were selected from
WINGS \citep{Fasano06}.

\subsection{PM2GC}

The PM2GC (Calvi et al. 2011)
is a database
built on the basis of the Millennium Galaxy Catalogue (MGC), a deep and
wide B-imaging survey along an equatorial strip of $\sim
38$deg$^{2}$ obtained with the INT (Isaac
Newton Telescope). The design, execution, reduction, object detection and
preliminary analysis of this survey are described in \citet{LI}.
The MGC field lies within the 2dFGRS Northern Galactic Cap region and the SDSS
region and a detailed comparison of the MGC
with these surveys is described in \citet{cross}.

We constructed the PM2GC catalogue restricting ourselves 
to galaxies brighter than $M_{B}<$-18.7 with a spectroscopic redshift
in the range 0.03$\leq$z$\leq$0.11 (3210 galaxies), taken from the
MGCz catalogue, the spectroscopic extension of the MGC that
has a $96\%$ spectroscopic completeness at these magnitudes \citep{DR}.


By applying a friends-of-friends algorithm we were able to identify a
catalogue of 176 galaxy groups with at least three members in the
redshift range 0.04$\lesssim$z$\lesssim$0.1 containing in total 1057
galaxies (PM2-G, hereafter {\it groups}). We consider members of the groups only those galaxies
that after several iterations are within
$\rm 1.5R_{200}$\footnote{$\rm R_{200}$
is the approximation of the virial radius computed as in \citet{FIN}.}
from the
group centre and $\rm 3\sigma$ (velocity dispersion)
from the group redshift. Galaxies that do
not satisfy the group linking
criteria adopted have been placed either in the catalogue of single
field galaxies (PM2-FS, hereafter {\it single}), that comprise the isolated galaxies, or in the
catalogue of binary field galaxies (PM2-FB, hereafter {\it binary}) which comprise
the systems with two galaxies within 1500 $\rm km \, s^{-1}$
and 0.5 $h^{-1}$ Mpc.
The redshift range of these catalogues is
0.03$\leq$z$\leq$0.11. All galaxies in the environments described above
and galaxies
excluded from the final virialized groups by the FoF procedure are
collected in the ''general field'' sample (PM2-GF, hereafter {\it general field}). The methods and
the presentation of catalogues are described in \citet{Ca} and the
samples are available online on the web page of the MNRAS paper and of the
MGC\footnote{http://www.eso.org/~jliske/mgc/}.

For the analysis discussed in this paper
we decided to limit the single, binary and general field catalogues
to the same redshift
range of groups (0.04$\lesssim$z$\lesssim$0.1) and for general field we also excluded
galaxies in group with edge problems. Moreover, in this paper
we considered as ``group'' galaxies
only members of groups with a velocity dispersion
$\sigma<500 \rm km \, s^{-1}$, to eliminate from our group sample a
possible contamination from clusters.

\subsection{WINGS}

Designed to investigate the properties of galaxies in clusters
and their connection with the cluster properties,
WINGS\footnote{http://web.oapd.inaf.it/wings} \citep{Fasano06}
is a multiwavelength survey based on deep optical (B,V) wide field
images ($\sim$ 35'$\times$35') of 76 clusters at 0.04$<$z$<$0.07. The
targets were selected in the X-ray from the ROSAT Brightest Cluster
Sample, and its extension \citep{eb2,eb3} in the northern hemisphere,
and the X-ray Brightest Abell-type Cluster sample \citep{eb1} in the
southern hemisphere, and span a wide range in velocity dispersion
($\sigma$ typically between 500-1100 km s$^{-1}$) and X-ray luminosity
(L$_{X}$ between 0.2-5$\times10^{44}$erg \, s$^{-1}$).

In addition to the optical imaging data a number of follow-ups
were carried out to obtain a large set of homogeneous informations for
galaxies in WINGS clusters. WINGS-SPE is the spectroscopic
survey conducted with the spectrographs WYFFOS@WHT and 2dF@AAT for a
subsample of 48 clusters for galaxies with a fiber aperture magnitude
 V$<$21.5 \citep{cava}.
In addition, near-infrared (J, K) observations of 28 clusters with
WFCAM@UKIRT \citep{VA} and U-band
imaging for a subsample with wide-field cameras at different
telescopes (INT, LBT, Bok, Omizzolo et al. in prep.)
have been obtained. An Omegacam/VST  U,B and V follow-up of about 50 WINGS
clusters is underway.

For our analysis, we have considered 21 of the 48 clusters with
spectroscopy. This is the subsample that provides a spectroscopic
completeness larger than 50\% (see Table1 in \citet{VU1}).
Only spectroscopically confirmed members within $\rm 0.6R_{200}$
(the largest radius generally covered in clusters)
will be considered. For our analysis, WINGS galaxies
were weighted for spectroscopic incompleteness
using the ratio between the number of galaxies with a spectroscopic redshift
and the number of galaxies in the parent photometric catalogue, as a function
of galaxy magnitude, as described in \citet{cava}.
A detailed description of redshift measurements, cluster
membership and completeness level is given in \citet{cava}.

\subsection{Morphological classification}

All galaxies in our samples have been morphologically classified
using MORPHOT, an automatic non parametric tool designed to obtain
morphological type estimates of large galaxy samples \citep{Fasano06,
F11}, which has been shown to be able to distinguish between
ellipticals and S0 galaxies with unprecedented accuracy. It combines a
set of 11 diagnostics, directly and easily computable from the
galaxy image and sensitive to some particular morphological
characteristic and/or feature of the galaxies, providing two
indipendent estimates of the morphological type based on: (i) a
Maximum Likelihood technique; (ii) a Neural Network
machine. The final morphological estimator combines the two techniques
and the comparison with visual classifications of SDSS images provides
an average difference in Hubble type $\Delta T$ ($\leq 0.4$) and a
scatter ($\leq 1.7$) comparable to those among visual classifications
of different experienced classifiers.

The classification process has been performed using B-band images for
PM2GC galaxies and V-band images for WINGS, after testing that no
significant systematic shift in broad morphological classification (ellipticals
E, lenticulars S0 or late-types LT)
exists between the V and B WINGS images
(see Calvi et al. 2012 for details).
In Table~\ref{t2} we list the morphological
fractions of elliptical, S0, early-type (ellipticals + S0s) and
late-type galaxies in each sample for galaxies with $\rm
log_{10}M_{\star}/M_{\sun} \geq 10.25$.

The morphological catalogue is available online
as Table~7 in the electronic version of the journal.
The different columns
indicate: (1) galaxy serial number in MGC; (2) MORPHOT classification,
see \citet{Fas12} for a detailed classification scheme. Here,
TypeMOR $<$-4.25
Ellipticals, -4.25$\leq$TypeMOR$\leq$0  S0s, TypeMOR$>$0 Late-types,
TypeMOR=99.0 for objects
that MORPHOT was not able to classify, mostly because suffering of
edge problems.
\begin{table}
\centering
\resizebox {0.48\textwidth }{!}{
\begin{tabular}{lcccc}
\hline\hline
& \multicolumn{4}{c}{\textbf{Galaxy type}}\\
\textbf{Envir.} & Ellipticals & S0s & Late-type & Early-type \\
\hline\hline
WINGS  & 33.8$\pm$1.5\% & 50.7$\pm$1.5\% & 15.4$\pm$1.0\% &
            84.5$\pm$1.0\%\\
gen.field &  27.0$\pm$1.3\% & 28.7$\pm$1.3\% & 44.3$\pm$1.5\% &
            55.7$\pm$1.5\%\\
groups & 31.8$\pm$2.4\% & 31.3$\pm$2.4\% & 36.9$\pm$2.5\% &
            63.0$\pm$2.5\%\\
binary & 25.3$\pm$3.5\% & 25.8$\pm$3.6\% & 48.8$\pm$4.0\% &
            51.1$\pm$4.0\%\\
single &  21.5$\pm$2.3\% & 24.2$\pm$2.5\% & 54.2$\pm$2.8\% &
            45.7$\pm$3.0\%\\
\hline
\end{tabular}}
\centering
\caption{Fractions of each morphological type
  in the PM2GC and WINGS mass-limited samples with
  $M_{\star}$=10$^{10.25}M_{\odot}$. WINGS = clusters (corrected for
completeness). Early-type galaxies comprise ellipticals and S0s. Errors are binomial. Data taken from \citet{Ca2} with the correction for groups which now comprises only groups with a velocity dispersion
$\sigma<500 \rm km \, s^{-1}$.
}\label{t2}
\end{table}

\section{The galaxy MF}

\subsection{Estimate of galaxy stellar masses and definition of the samples}
As argued by \citet{BJ} the galaxy stellar M/L ratio is a function of color
according to the relation
\begin{equation}\label{eq1}
\log_{10}(M_{\star}/L)=a_{\lambda}+b_{\lambda}Color
\end{equation}
which is robust to uncertainties in stellar populations and
galaxy evolution modeling, including the effects of modest
bursts of recent star formation.

As described in \citet{Ca} using (\ref{eq1}) we derived the stellar masses
for PM2GC and WINGS galaxies considering the rest frame $(B-V)$ color,
computed from the SDSS Galactic extinction-corrected model magnitudes
 in $g$ and $r$ (for the PM2GC) and the observed B and V WINGS magnitudes,
with $a_{B}$=-0.51 and $b_{B}$=1.45 for the Bruzual \& Charlot model,
solar metallicity and
a \citet{Sa} IMF (0.1-125 M$_{\odot}$).
Subsequently
the masses have been scaled
to a \citet{Kr} IMF applying a conversion factor and then compared
with other estimates
obtained with different methods to verify the absence of offsets.
For PM2GC galaxies we compared with
masses obtained using SDSS-CAS magnitudes and with SDSS DR7 masses
(see \citet{Ca} for details),
while for WINGS galaxies the masses have been compared with those
determined by the \citet{fr} spectro-photometric model and with
the Sloan \citet{VU1,fr}. In all cases the agreement
is satisfactory, with no systematic offset and an rms scatter
of 0.1-0.2 dex. The uncertainty on galaxy mass estimates is 0.2-0.3dex,
as described in the papers listed above.

Since our goal is to perform a detailed analysis of the galaxy
properties, we restricted to a galaxy sample complete in mass so that
all types of galaxies are potentially observable above this mass. The
galaxy stellar mass completeness limit was computed as the mass of the
reddest galaxy at the upper redshift limit. For PM2GC the reddest
color corresponding to $B=20$ is $B-V=0.9$ at our
redshift upper limit $z=0.1$, and the mass limit is equal to
$M_{\star}=10^{10.25}M_{\odot}$. The mass limit for WINGS is lower
($M_{\star}=10^{9.8}M_{\odot}$), but for homogeneity we
adopted the same galaxy mass limit as for the PM2GC.

The number of galaxies in the
various samples above our mass limit are
listed in Table \ref{tt}.
\begin{table}
\centering
\begin{tabular}{cccc}
\hline
\hline
\textbf{\textit{Red. range}} & \textbf{\textit{Environment}} & \multicolumn{2}{c}{\textbf{\textit{Num. of galaxies}}}\\

\hline
\hline
0.04$<$ z$<$0.07 & clusters  & 690 (1056) \\

0.04$\lesssim$ z $\lesssim$ 0.1 & general field & 1188 \\
''& groups & 409 \\
''& binary & 174 \\
''& single & 334 \\
\hline
\end{tabular}
\centering
\caption{Number of galaxies
  in the PM2GC and WINGS mass-limited samples with
  $M_{\star}\geq 10^{10.25}M_{\odot}$. 
  The WINGS number between
  brackets is weighted for spectroscopic incompleteness.
}\label{tt}
\end{table}

\subsection{The method}

The analysis of the galaxy MF that follows is based on a visual
ispection of the shape of the data distribution, on the application
of the Kolmogorov-Smirnov (K-S) test, and on the analysis of the
parameters of Schechter function fits.
We built the histograms of the mass distributions of galaxies setting
the bin width at 0.2 dex while the number density of galaxies is
obtained summing all galaxies in each bin.
For a better visual comparison between the different
galaxy samples in the plots, since we are interested mainly in the shape of the
mass distributions and not in the counts, we normalized each mass
function to the integrated stellar mass above the completeness limit, unless
otherwise stated.
In all plots, the errors along the $\rm x$ direction represent
the bin size
while in the $\rm y$ direction they are computed as poissonian errors as
in \citet{GEH}.

In combination with these graphs, a ``low probability'' ($P_{K-S}<5\%$)
of the K-S test is a statistically
significant result to assess that two samples are different; on
the contrary a ``high probability'' does not prove that they are drawn
from the same distribution but only that the test is unable to find
differences.

In addition, we perform  \cite{Schechter76} fits of the mass
functions using the least square fitting method.
With this formalism
the galaxy stellar MF can be described as
\begin{equation}\label{eq:sc}
\Phi (M) = (\ln 10) \times \Phi^* \times [10^{(M-M^*)(1+\alpha)}] \times
exp[-10^{(M-M^*)}]
\end{equation}
where $M = \log (M_{\star}/M_\odot)$, $\alpha$ is the low-mass-end slope,
$M^*$ 
is the logarithm of the characteristic stellar mass at which
the MF exhibits a rapid change in the slope, and $\Phi^*$  is the
normalization.

\subsection{Comparison with previous works}

\begin{figure}
    \vspace{-10pt}
	\includegraphics[scale=0.42]{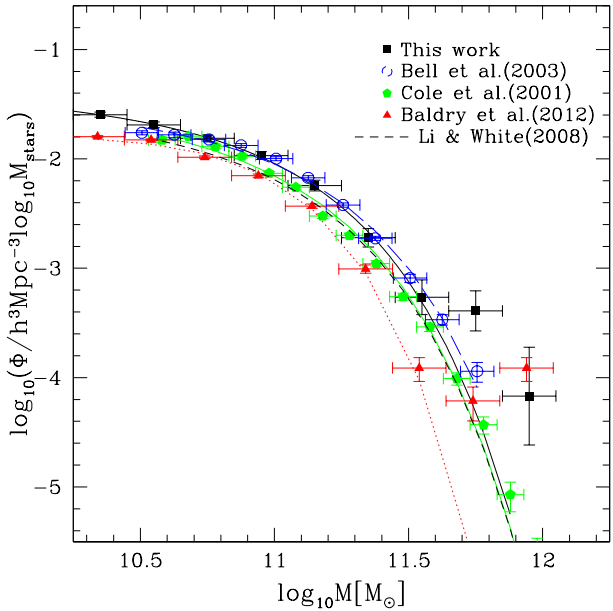}
\caption{Comparison between our general field MF and literature
results. Masses are in units of $M_{\odot}$ for a Kroupa
IMF. $\Phi$ values are in units of
number per $h^{-3} \, \rm Mpc^{3}$ per decade of mass ($dex^{-1}$).
Best fit Schechter functions are shown as lines.
}
\label{F0}
\end{figure}

First of all we compare our MF with previous results from
the literature to check if they are in agreement.
Fig. \ref{F0} shows the comparison between the MFs
of the PM2GC general field
and the MFs from 2dFGRS-2MASS (Table 4 in \citealt{CO}),
SDSS-2MASS (Table 5 in \citealt{BE}), SDSS-DR7 \citealt{Li}
and the recent GAMA result from \citealt{BA4}. All the MFs in this plot
are given in units of
number per $h^{-3} \, \rm Mpc^{3}$ per decade of mass ($dex^{-1}$).
Masses in $M_{\odot}$ are all converted to a Kroupa IMF.
For our work, Cole's, Bell's and Baldry's we plot the binned
MFs (symbols) and the best-fitting Schechter functions, while
for Li \& White's we can only show the Schechter fit they
provide (the black short dashed line in the plot).

The shape of our MF is in very good agreement with all previous estimates.
As for the absolute normalization, the only MF that tends to be slightly
lower is the one from \citet{BA4}.
The excess of very massive galaxies
($\rm log_{10}M_{\star}/M_{\odot} > 11.7$) with respect to the Schechter
function is similar in our sample and in GAMA, and is present
in many previous studies (e.g. \citealt{PAN,Li}).

The agreement among the MFs is confirmed by the Schechter parameters.
Our $\rm log_{10}M_{\star}/M_{\odot} =10.96 \pm 0.06$,
$\alpha = -1.1 \pm0.1$ and ${\phi}^{\star} =0.011\pm0.004$
are fully consistent with
Cole's $\rm log_{10}M_{\star}/M_{\odot} =10.97 \pm 0.01$,
$\alpha = -1.18 \pm0.03$ and ${\phi}^{\star} =0.009\pm0.0014$, Li's
$\rm log_{10}M_{\star}/M_{\odot} =10.85 \pm 0.53$,
$\alpha = -1.155 \pm 0.008$ and ${\phi}^{\star} =0.0083\pm0.0002$, and Bell's
$\rm log_{10}M_{\star}/M_{\odot} =11.02 \pm 0.02$,
$\alpha = -1.10 \pm0.02$ and ${\phi}^{\star} =0.0102\pm0.0005$
when they are all converted to our units.
Baldry et al. fit their data with a double Schechter therefore
parameters cannot be compared.

\section{Results: the MF in different environments}

In this section we focus our attention on
how the galaxy MF changes with galaxy ``global
environment'' in the local Universe comparing PM2GC and WINGS
above our completeness limit of $M_{\star}=10^{10.25}M_{\odot}$. An analysis
of the WINGS's MF down to $M_{\star}=10^{9.8}M_{\odot}$
can be found in \citet{VU1}.

As the presence of Brightest Cluster Galaxies (BCGs), defined as the
single brightest galaxy in each galaxy cluster, could alter the total mass
distribution we investigated the MFs also excluding the
BCGs in the WINGS sample.
We remind the reader that, in the case of WINGS, each
galaxy is weighted by its correction for spectroscopic incompleteness.

\subsection{General field versus clusters}

\begin{figure}
    \vspace{-10pt}
	\includegraphics[scale=0.42]{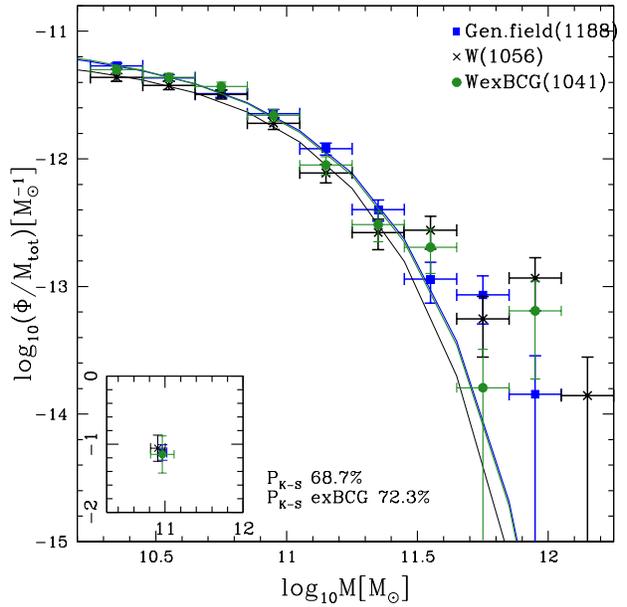}
\caption{Comparison between the mass distribution of galaxies of all
morphological types in general field (blue empty squares) and in WINGS with
(black crosses) and without (green filled circles) the BCG
galaxies. MFs are normalized using the total integrated stellar mass,
above the mass completeness limit. Numbers in the brackets are the total number of
galaxies in each sample observed above the completeness limit, for WINGS they
are weighted for incompleteness. The
relative K-S probabilities are also shown. Errors on the y axis are
poissonian. The inset shows
the $\rm \alpha$ and $\rm log_{10}M^{\star}$
of each sample.}
\label{F1}
\end{figure}

Fig. \ref{F1} shows the comparison between the mass distribution of
galaxies in the general field and the
mass distribution of all cluster galaxies.

Looking at the plot, the overall similarity of the shape of the mass
functions of clusters and general field is rather striking.
For galaxies
with masses up to $\rm log_{10}M_{\star}/M_{\odot} \sim 11.5$ the MFs
overlap, while in at least two of the four most massive bins at $M_{\star} >
10^{11.5}M_{\odot}$  the WONGS sample exhibits an excess of galaxies
compared to the PM2GC. This excess is due to the presence of BCG
galaxies. Removing the BCGs, the MFs of general field and clusters
 are similar within the errors at all masses.
The K-S test also
finds no difference both when we include the BGCs ($P_{K-S}\sim 69\%$)
or not ($P_{K-S}^{exBCG}\sim 72\%$).

The similarity of the cluster and general field MFs is also
confirmed by the Schechter fits shown in Fig. \ref{F1} and by the
analysis of the best fit parameters, that are similar
(inset in Fig. \ref{F1} and Table~\ref{tt2}).\footnote{In Table~\ref{tt2},
for the PM2GC, we give the $\rm \Phi^{\star}$
in units of number per $h^{-3} \, \rm Mpc^{3}$ per decade of mass
($dex^{-1}$), while for clusters we can't obtain a similar estimate because
of the uncertainty in assessing the volume relative to galaxies
that are considered cluster members.}
Indeed, taking into account the fact that the two parameters are
correlated, we explored a grid of $\alpha$ and $M_{\star}$ parameters,
finding the corresponding $\chi ^{2}$ values and the likelihood of
having the same couple of values and found that MFs are in
agreement within 1$\sigma$. We note that, as seen in previous works (e.g.
\citealt{PAN,Li,BA4}), the
Schechter function is unable to fit the very massive end.


The general field sample is the sum of group galaxies, (which dominate the
general field, and whose MF is very similar to the general field mass
function, for its shape, Schechter fit and K-S test),
binary system galaxies, single galaxies and galaxies
that, although located in a trial group, did not make it into the
final group sample. In the following section our aim is to understand
if differences in the galaxy MF become appreciable
when considering these finer division of environments.

\begin{table*}
\centering\resizebox {0.55\textwidth }{!}{
\begin{tabular}{|l|c|c|c|}
\hline
\hline
 & \multicolumn{3}{c}{\textbf{\textit{Schechter parameters}}}\\
\textbf{\textit{Environments}} & $\alpha$&$\rm log_{10}M^{\star}$ &$\phi^{\star}(h^3 \, \rm Mpc^{-3} \,log_{10}(M^{-1}))$ \\
\hline
\hline
WINGS & $-1.1\pm 0.3$ & $10.96\pm 0.15$ & ... \\
WINGSexBCG & $-1.1\pm 0.2$ & $10.90\pm 0.09$  & ... \\
general field &$-1.1\pm 0.1$&$10.96\pm 0.06$ & $(1.1\pm 0.4)\times 10^{-2}$ \\
groups &$-1.2\pm 0.1$ &$11.05\pm 0.08$ &$(0.3\pm 0.1)\times 10^{-2}$ \\
single & $-1.3\pm 0.1$&$10.94\pm 0.04$ & $(0.3\pm 0.1)\times 10^{-2}$\\
binary & $-0.6\pm0.5$& $10.75\pm0.23$ & $(0.3\pm 0.1)\times 10^{-2}$ \\
\hline
\end{tabular}}
\centering
\caption{Best-fit Schechter parameters for the different samples.}\label{tt2}
\end{table*}


\subsection{The MF in groups, binaries, singles and clusters}

\begin{figure*}
    \vspace{-10pt}
    \includegraphics[scale=0.7]{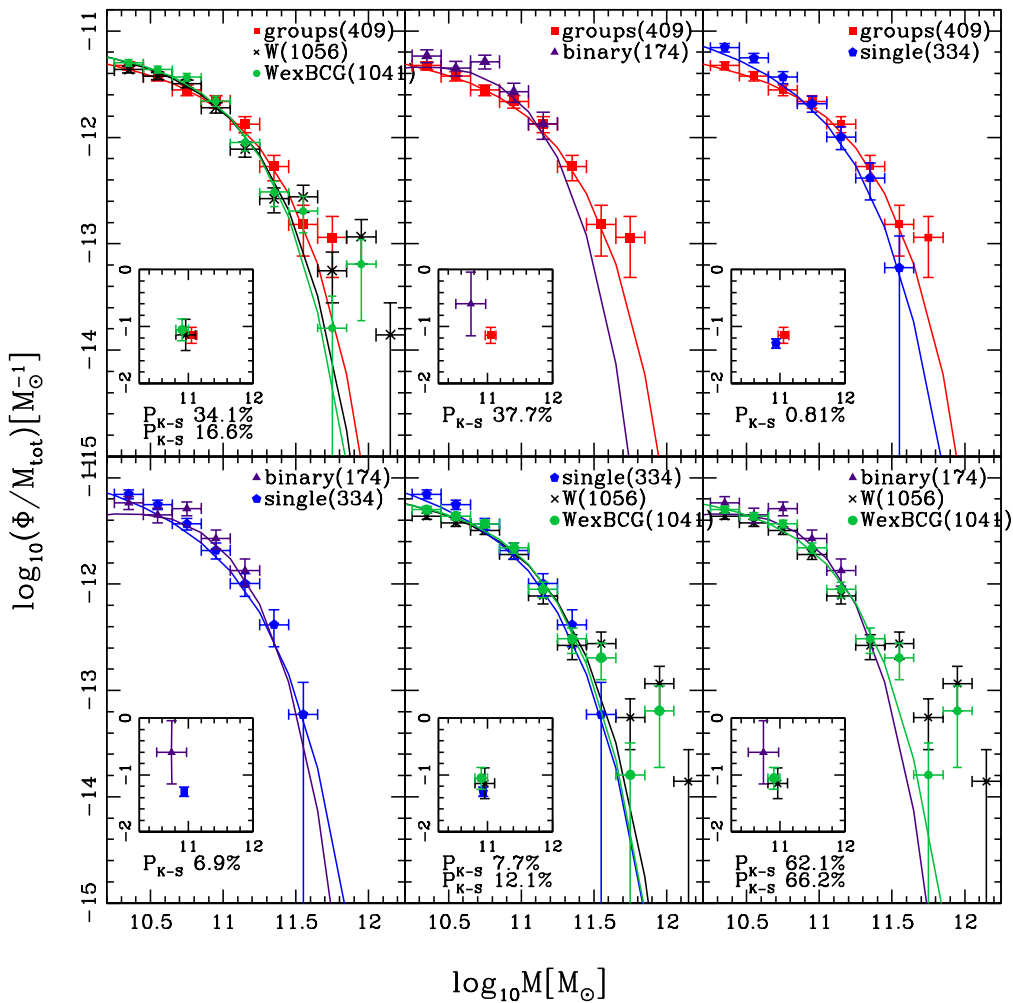}
    \caption{Comparison of the mass distributions of galaxies of all
morphological types, for the mass limited samples, for different pairs
of environments: groups (red squares)
vs WINGS (top left panel) with (black crosses)
and without BCGs (green circles), groups vs binary (filled dark triangles,
top central panel),
groups vs single (blue filled pentagons, top right panel),
single vs binary (bottom left
panel), WINGS, with (black crosses) and without BCGs (green circles) vs
single (bottom central panel), WINGS, with and without
BCGs vs binary (bottom right panel). Mass distributions
are normalized using the total integrated stellar mass, above the mass completeness limit.
For goups we considered only groups with $\sigma <$ 500 $\rm km s^{-1}$. Errors are
poissonian in the \textit{y} direction and equal to the bin size in the
\textit{x} direction. Numbers in the brackets are the total number of
galaxies above the completeness limits. In the bottom left corner of the
panels we show the relative K-S probability.}
\label{F2}
\end{figure*}

Fig. \ref{F2} shows the mass distribution of galaxies comparing different
pairs of environments. Also in this case we use the WINGS
sample both with and without BCGs.
The fact that we consider as group galaxies only
those galaxies which are in groups with a velocity dispersion less
than 500 $\rm km \, s^{-1}$ makes us confident that our findings
for groups are not influenced by galaxies in structures as massive
as WINGS's, therefore the group and cluster distributions sample
truly different environments.

The shapes of the MFs of groups and clusters
show a rather similar trend, as expected given the results shown above
and the similarity of the general field and group MF.
At masses $\rm
log_{10}M_{\star}/M_{\odot}<11.2$ the distributions overlap, while in
two intermediate mass bins
the distribution in groups tends to be higher
than the cluster one (top left panel).  The K-S test is
not able to reveal difference being $P_{K-S}\sim 34\%$ and
$P_{K-S}^{exBCG}\sim 16\%$ with and without BCGs, respectively.
The compatibility of cluster and group MFs is also
confirmed by the Schechter fit parameters (Table~\ref{tt2}).
In addition, we note that both in clusters and groups
galaxies more massive than
$\rm log_{10}M_{\star}/M_{\odot}>11.5$ are found, but the cluster
MF extends to even higher masses than the groups when including
BCGs.


Considering the MF of the single, that is the extreme
``low-mass halo'' environment, we note that the slope of the single
MF in Fig.\ref{F2} appears steeper than any other
environment (groups, binaries and clusters).  The K-S test is able to
conclude that the MFs of group and isolated galaxies are
statistically different ($P_{K-S}< 1\%$). Quite low K-S probabilities
(of the order of 7-8$\%$) are also suggesting that the visual
differences between the MF of single galaxies and that of
clusters and binaries might be real.  Indeed, single galaxies have the
steepest value of $\alpha$ in the Schechter fit, although the
differences with the other environments are not statistically robust
given the errorbars.

For binary galaxies it is difficult to quantify the differences given
the low number statistics and the fact that the upper mass of binary
is lower than the others, as discussed in the next section.  From the
slope of the binary MF there may be a hint that this
is flatter than others at low masses, but no statistically robust
difference can be found based on the KS test, and the Schechter
parameters are 
unconstrained for binaries.

To conclude, no statistically significant difference has been found
between the galaxy MF in groups and clusters,
while a variation with global environment starts to be
appreciable when considering single galaxies, that show a steeper
MF, therefore are proportionally richer in lower-mass galaxies, than
other environments.  It is worth noting that single galaxies
represent less than a third (28\%)
of the general field population above our mass limit, as can be
inferred from Table~\ref{tt}, therefore their influence on the total general field
MF is small.

\subsection{Cut-off in mass}

In addition to the similarities and differences described above, it is
interesting to observe that in binary and single systems there are no
galaxies with masses $M_{\star}\geq 10^{11.2}M_{\odot}$ and
$M_{\star}\geq 10^{11.55}M_{\odot}$, respectively, while in groups and
clusters there are galaxies up to $M_{\star}\sim 10^{11.75}M_{\odot}$
and $M_{\star}\sim 10^{12}M_{\odot}$, respectively,
even excluding cluster BCGs.
This might suggest that galaxies in different environments could reach
different upper masses.

To better quantify the differences, in Table
\ref{t4} we show the number of galaxies in each environment above and
below $M_{\star}=10^{11.2}M_{\odot}$ (the upper limit of masses
for binary system galaxies), and their number ratio.
The ratio varies with environment being higher
in clusters, than groups, than single and binary galaxies.

\begin{table}
\centering
\resizebox {0.51\textwidth }{!}{
\begin{tabular}{cccc}
\hline
\hline
\textbf{environment} & \textbf{\textit{$M_{\star}\geq 10^{11.2}M_{\odot}$}} & \textbf{\textit{$M_{\star}< 10^{11.2}M_{\odot}$}} & \textbf{$M_{(\geq 11.2)}/M_{(< 11.2)}$} \\
\hline
\hline
WINGS(exBCGs) & 46(34) & 644 & $7.1\pm 1.1\%(5.2\pm 0.9\%)$\\
general field & 52 & 1136 & $4.6\pm 0.6\%$\\
groups & 23 & 386 & $6.0\pm 1.2\%$\\
binary & 0 & 174 & $0.0\pm 1.1\%$\\
single & 12 & 322 & $3.7\pm 1.1\%$\\
\hline
\end{tabular}}
\centering
\caption{Number of galaxies
  in the PM2GC and WINGS mass-limited samples with mass
  $M_{\star}\geq 10^{11.2}M_{\odot}$ and $M_{\star}< 10^{11.2}M_{\odot}$ and their ratio as a
percentage.}\label{t4}
\end{table}

In order to assess the significance of the variation of the upper mass
limit we performed a Montecarlo simulation to understand whether this
effect could be due to low number statistics in the least massive
environments.

Using each time 1000 simulations, we
extracted from the group sample the same number of galaxies once as
in the single sample and once as in the binary sample, and then
extracted from the single sample the same number of galaxies as in
binary systems.

\begin{table}
\centering
\begin{tabular}{lcc}
\hline\hline
& Fraction & \textbf{Median upper mass($\rm M_{\star}$)}\\
\hline\hline
groups-bin & 0.3\% &$10^{11.7}$\\
groups-sin & 1.5\% &$10^{11.8}$\\
sin-bin &  1.6\% & $10^{11.4}$\\
\hline
\end{tabular}
\caption{Fractions of simulations which reach an upper mass at least as low
as the observed mass +0.1dex
and values of
the median upper mass reached
in the Montecarlo simulations comparing
group and binary, group and single, single and binary samples.}
\label{cut}
\end{table}

In Table~\ref{cut} we show the median upper masses for the different
simulations: they are always significantly higher than the cut-off
mass observed in singles and binaries.

Table~\ref{cut}  also gives
the percentages of simulations that display a cut-off mass
equal or lower than the observed mass + 0.1dex. The 0.1dex is added
to take into account the errors on the masses.
These percentages can be seen
as the probability that the low mass cut-off observed in binary and
single systems is due to the small number of galaxies in the
sample. This probability is always very low (e.g. 0.3\%, that is 3
out of the 1000 simulations).
This
demonstrates that even reducing the number of galaxies in groups
to the same numbers
of singles and binaries, we would
expect to observe massive galaxies in groups.
We thus
conclude that effectively the cut-off mass varies with environment,
and that massive galaxies can only be found in environments that
correspond to more massive dark matter haloes, as also shown by previous
works (e.g. \citealt{Yang} and expected from simulations \citealt{Mo}).




To summarize, at low redshift, (1) no significant difference can be
found in the MFs of the general field, clusters and groups by
analyzing the parameters of the Schechter fits or the KS test.
Our results resemble those at higher redshifts of \citep{VU3},
that found a similar MF in clusters, groups and general
field at $z=0.4-0.8$.
(2) For binary systems, the KS test is inconclusive and the Schechter
parameters are unconstrained, hence secure conclusions cannot be
reached for this type of environment.  (3) Differences have been found
in the MF of single galaxies compared to other
environments. The MF of singles appears steeper
than the others.  According to the K-S test, the difference is
statistically significant between single and group galaxies, while it
is only marginally hinted by the K-S values when comparing with the
binary and cluster samples.  Schechter fit parameters are unable to
detect statistically significant differences.
(4) Very
massive galaxies are only found in the most massive environments,
groups and clusters, while they are absent among single and binary
galaxies. The presence or absence of such massive galaxies is unable
to affect both the Schechter parameters (being $M^{\star}$ always
lower than the mass of these galaxies) and the K-S test (dominated by the most
frequent, lower mass galaxies and unaffected by the few very massive
ones).  Only a separate analysis of the mass cut-off has been able to
highlight the dependence of the cut-off mass on global environment.


\section{The galaxy MF by morphological type}

In the previous sections we found the somewhat unexpected result that
the MF is similar in different global environments, except
when analyzing single galaxies separately and when studying in detail
the cut-off mass.

Now we attempt to examine the MF of different morphological
types, to address two main questions: how the MF differs
from a galaxy type to the other, and whether the MF of each
given type varies with environment.

\subsection{The galaxy MF of different morphological types in
each given environment}

\begin{figure*}
    \vspace{-5pt}
    \includegraphics[scale=0.8]{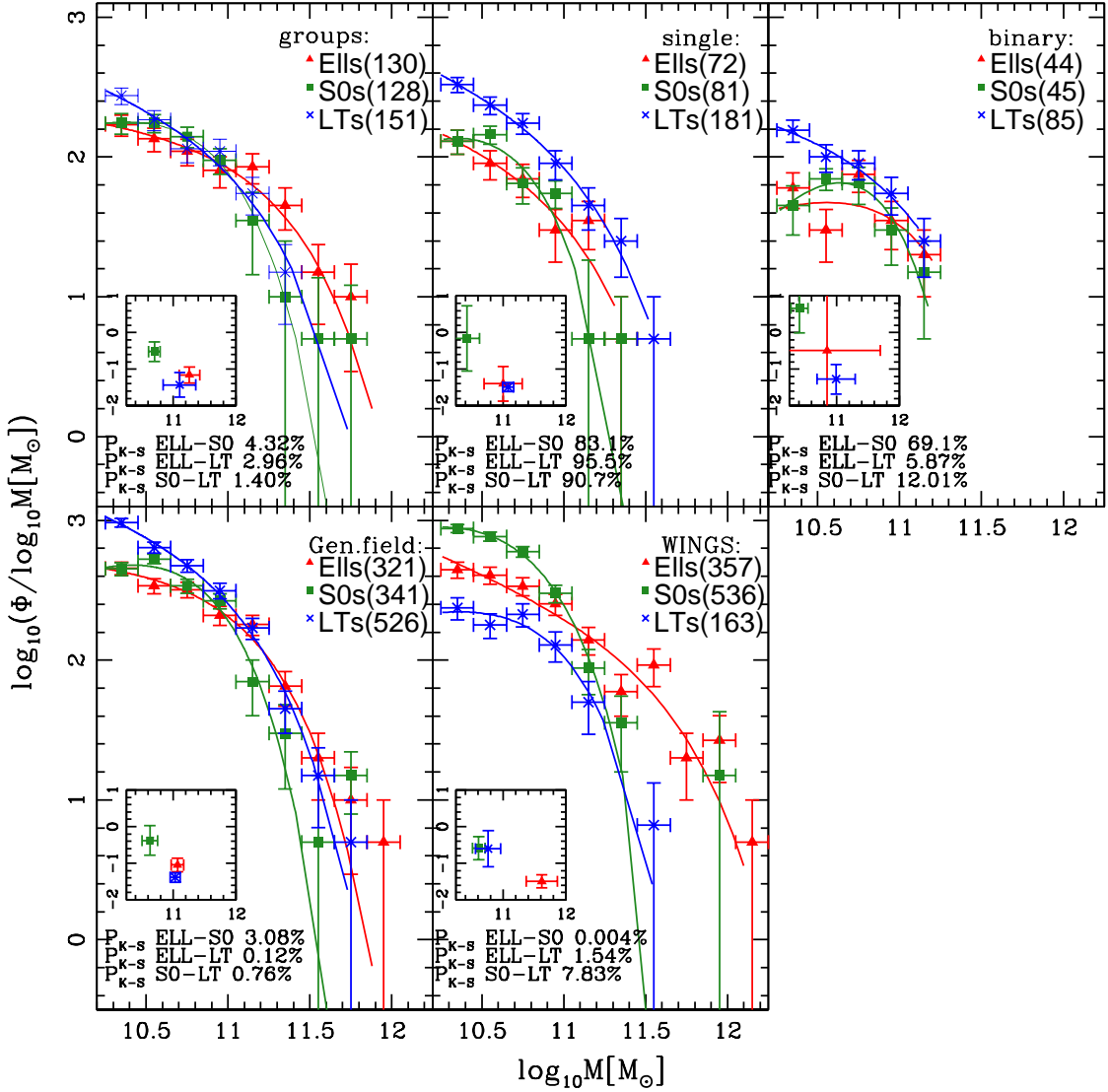}
    \caption{Mass distribution of galaxies in the groups (top left
panel), single (top middle panel), binary (top right panel), general field
(bottom left panel), clusters (bottom right panel). The K-S
probabilities are shown. Red triangles are elliptical galaxies, green
squares S0s and blue crosses late-type galaxies. Lines represent
Schechter fits.
 Errors are poissonian
errors in the \textit{y} direction and are equal to the bin size in the
\textit{x} direction. Numbers in brackets are the number of galaxies in each
morphological class, above the respective mass limit and are weighted
for WINGS.}
\label{F3}
\end{figure*}

We start analyzing the MFs of different morphological types
in each given environment.  In this case, we don't apply any
normalization to the MFs, to show which morphological type
dominates in number as a function of mass.
Table~\ref{sc_par} gives the best fit Schechter parameters and Fig.~\ref{F3} shows the
mass distributions of galaxies in each environment.


In the general field (bottom left panel),
the total MF is dominated by
late-type galaxies at low masses ($\rm log M_{\star}/M_{\odot}\lesssim
11$), and by a mix of late-types and ellipticals at higher
masses. Instead, in the single and binary systems (central and right
upper panels), it is dominated
by late-type galaxies at all masses.
In groups (upper left panel), the
most numerous types at masses $\rm \lesssim 11$ are late-types and
S0s, except in the first mass bin ($\rm 10.25- 10.5$) where late-type
galaxies are more numerous.  At higher masses, elliptical galaxies
dominate in groups.

The cluster environment (bottom right panel)
stands out for its peculiarity. Unlike
the other environments, its main population at masses $\rm \lesssim 11$
is composed of S0 galaxies but, as mass increases, the S0 MF sharply
decreases and ellipticals start to dominate.  The number of late-type
galaxies is steadily lower than S0s and ellipticals at all
masses.

From Fig.~\ref{F3} it is clear that the shape of the MF depends
on the morphological type in most environments. This is
confirmed by the K-S test, which finds incompatible distributions for
ellipticals, S0s and late-types in the general field, in groups and in
clusters, with the exception of cluster S0s and late-types whose MFs
are less distiguishable. These conclusions are generally confirmed by the
Schechter fit parameters shown in the insets of Fig.~\ref{F3}.

For binary system galaxies, the K-S test and the analysis of the
Schechter fits are inconclusive due to low number statistics, but in
the plot the late-type MF appears to be significantly different from
the MFs of the other types ($P_{K-S} = 5.9$ and 12.0 for ellipticals-late-types and
S0-late-types, respectively).

For single galaxies, the shape of the MF varies little between
late-type galaxies and ellipticals, and may differ for S0s, as
suggested also by the Schechter fit parameters. The K-S test is always
inconclusive.

We note from Table~6 that in all environments the S0's $\rm M^{\star}$ value is
significantly lower than those of ellipticals and late-types, except
in clusters where the late-type $\rm M^{\star}$ is almost as low as
that of S0s. The lowest S0 $\rm M^{\star}$ value is reached in the
single and binary samples.  The $\rm M^{\star}$ values of ellipticals
and late-type galaxies are similar in all environments, except
in clusters where the elliptical $\rm M^{\star}$ is significantly
higher.

We conclude that in general field the shape of the MF changes
from one morphological type to another, but in a way that depends on
global environment. The cluster morphological MFs have a
peculiar behaviour, both for the MF shape of the various
types and their relative numbers.  In the next section we examine in
detail the MF of each morphological type in different
environments.




\subsection{The MF of ellipticals, S0s, early-type and late-type galaxies in the general field versus clusters}

\begin{figure*}
    \vspace{-20pt}
    \includegraphics[scale=0.55]{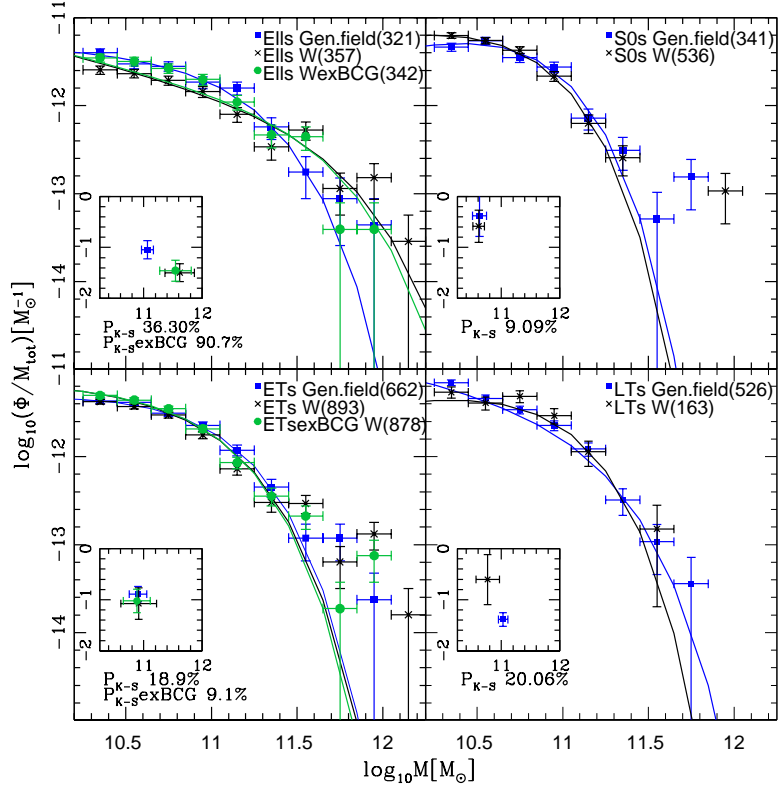}
    \caption{Comparison of the mass distribution of general field and
WINGS for elliptical galaxies (top left panel), S0 galaxies (top right
panel), early-type galaxies (bottom left panel) and late-type galaxies
(bottom right panel). Errors are defined as poissonian errors in the
\textit{y} direction and equal to the bin size in the \textit{x}
direction. Numbers in the brackets are the number of galaxies of each
type above the mass limit (for WINGS the number is weighted for
incompleteness). The K-S probabilities are also shown in the bottom
left corner. Mass distributions are normalized using the total
integrated stellar mass, above the mass completeness limit.}
\label{F8}
\end{figure*}

The comparison between the mass distribution of elliptical, S0, early
and late-type galaxies of general field and clusters is shown in Fig.
\ref{F8}.

In all cases, the K-S test is unable to detect any significant
difference between general field and clusters.  The analysis of the
Schechter parameters (Table~6) and the inspection of the plot,
instead, reveal a few differences.

Ellipticals in clusters, even when excluding the BCGs, have a higher
$M_{\star}$ and a lower value of $\alpha$ than ellipticals in the
general field. This is due to the excess of ellipticals with
masses $log M_{\star}/M_{\odot}>11.5$ in clusters compared to the field visible in the top
left panel of Fig.~\ref{F8}.

The Schechter parameters for S0s in clusters and general field are
instead statistically indistinguishable.  We note that, given the
large errorbars, the S0 Schechter $\alpha$ is essentially
unconstrained, and a visual inspection of the plot may suggest a
steeper low-mass end in clusters.


When ellipticals and S0s are considered together, the
early-type MF is similar in clusters and in the general field, and
also the Schechter parameters are compatible.  The environmental
variation of the MF of ellipticals seen in the top left panel gets
diluted when adding them up with lenticulars, and no significant
difference with environment is left when considering all early-type
galaxies.  Looking at the numbers in the plots, one can notice that
the general field consists of a similar number of ellipticals and S0s, while
WINGS clusters are dominated by S0s. The morphological fractions are
given in Table~1, and a detailed study of the variation of the
morphological mix with environment can be found in \citet{Ca}.

Coming to late-type galaxies, at low masses the shape of their mass
function in WINGS is slightly flatter than in general field. As also
indicated by the Schechter $\alpha$ parameter, there is a small
relative deficit of low-mass late-type galaxies in clusters compared
to the general field.

In conclusion, the only variations we are able to detect are an
excess of massive ellipticals and a small deficit of low mass late-type
galaxies in clusters compared to the general field. We cannot exclude
that, with better statistics, environmental variations of the S0 MF
could be found.

\begin{table*}
\centering\resizebox {0.75\textwidth }{!}{
\begin{tabular}{|l|l|c|c|c|c|c|c|c|} \hline
\multicolumn{5}{|c|}\centering{\textbf{\textit{Schechter parameters}}}\\
Environment&morph. type&$\alpha$&$\rm log_{10}M^{\star}$&$\phi^{\star}$  &$\phi^{\star}$ \\
& &&$M_{\odot}$& &$\rm h^{3}Mpc^{-3}log_{10}(M^{-1})$\\ \hline
\multirow{6}{*}{WINGS} & E & -1.5$\pm$0.2  & 11.62$\pm$    0.25 & 53.2 $\pm$  39.4      &.....\\
& EnoBCG & -1.4$\pm$    0.2 & 11.54 $\pm$   0.27 &64.9 $\pm$  51.6 &.....\\
& S0 & -0.6 $\pm$   0.3  & 10.61 $\pm$   0.10 & 849.8$\pm$  154.2&..... \\
& late & -0.6 $\pm$   0.5  & 10.77  $\pm$  0.20 &206.6 $\pm$  78.2 &.....\\
& early & -1.1 $\pm$   0.3  &10.91  $\pm$  0.16 & 712.2$\pm$  332.8&..... \\
& earlynoBCG & -1.0  $\pm$  0.2  & 10.89 $\pm$   0.11 & 775.1$\pm$ 248.0&.....\\
\hline
\multirow{4}{*}{general field} & E & -1.0 $\pm$   0.2 & 11.06  $\pm$  0.10 & 208.9$\pm$   61.2  &(2.8 $\pm$  0.8) $\rm \times 10^{-3}$\\
& S0 &  -0.4 $\pm$   0.4  & 10.63 $\pm$   0.12 &  520.4$\pm$   93.1  &(7.1 $\pm$  1.2)$\rm \times 10^{-3}$\\
& late & -1.4 $\pm$   0.1  & 11.03 $\pm$   0.08 &  271.9$\pm$   77.0  &(3.7$\pm$  1.0) $\rm \times 10^{-3}$  \\
& early &  -0.9 $\pm$   0.1 &10.90 $\pm$   0.07 & 612.8$\pm$  112.0   &(8.3$\pm$ 1.5)$\rm \times 10^{-3}$\\ \hline
\multirow{4}{*}{groups} & E & -1.1 $\pm$   0.2 & 11.26  $\pm$  0.16 & 57.4$\pm$   27.8    &(0.8 $\pm$  0.4)$\rm \times 10^{-3}$\\
& S0 & -0.5 $\pm$   0.3  & 10.70  $\pm$  0.09  &  178.0$\pm$   29.1  &(2.4 $\pm$ 0.4)$\rm \times 10^{-3}$\\
& late & -1.4  $\pm$  0.3  & 11.10 $\pm$   0.26 &  65.0$\pm$   56.7  &(0.9 $\pm$  0.8) $\rm \times 10^{-3}$\\
& early &-1.1 $\pm$   0.1  &11.09  $\pm$  0.09 &  156.8$\pm$  42.1  &(2.1 $\pm$ 0.6)$\rm \times 10^{-3}$\\ \hline
\multirow{4}{*}{binary} & E & -0.5  $\pm$  1.7  &10.85 $\pm$   0.85 & 48.2$\pm$   62.1 &(0.6 $\pm$ 0.8) $\rm \times 10^{-3}$\\
& S0 & 0.7  $\pm$  0.7  & 10.41 $\pm$   0.14 &    64.1 $\pm$ 18.3   &(0.9  $\pm$ 0.2)$\rm \times 10^{-3}$\\
& late &-1.3 $\pm$   0.4 & 11.00 $\pm$   0.30&  53.6$\pm$   47.4   &(0.7  $\pm$  0.6)$\rm \times 10^{-3}$\\
& early & 0.1 $\pm$   0.8  &10.58 $\pm$   0.21   &  137.7$\pm$   17.4   &(1.9  $\pm$  0.2)$\rm \times 10^{-3}$\\ \hline
\multirow{4}{*}{single} & E &   -1.4 $\pm$   0.5 & 11.00  $\pm$  0.30 & 38.2$\pm$  39.9    &(0.5 $\pm$ 0.5) $\rm \times 10^{-3}$  \\
& s0 & -0.2   $\pm$ 0.9  & 10.42  $\pm$  0.20& 158.2$\pm$  27.8    &(2.1  $\pm$ 0.4)$\rm \times 10^{-3}$\\
& late &-1.5 $\pm$   0.1 & 11.08  $\pm$  0.09 &   76.4 $\pm$  23.7  &(1.0  $\pm$  0.3)$\rm \times 10^{-3}$ \\
& early & -1.1 $\pm$   0.2  & 10.82 $\pm$   0.10& 147.7$\pm$   42.3   &(2.0  $\pm$  0.6)$\rm \times 10^{-3}$\\ \hline
\end{tabular}}
\caption{Best fit Schechter parameters for PM2GC groups, binary systems, single, general field and WINGS clusters.}
\label{sc_par}
\end{table*}

\subsection{The shape of the galaxy MF of each morphological type
in different environments}

\begin{figure*}
	\centering
    {\includegraphics[scale=0.52]{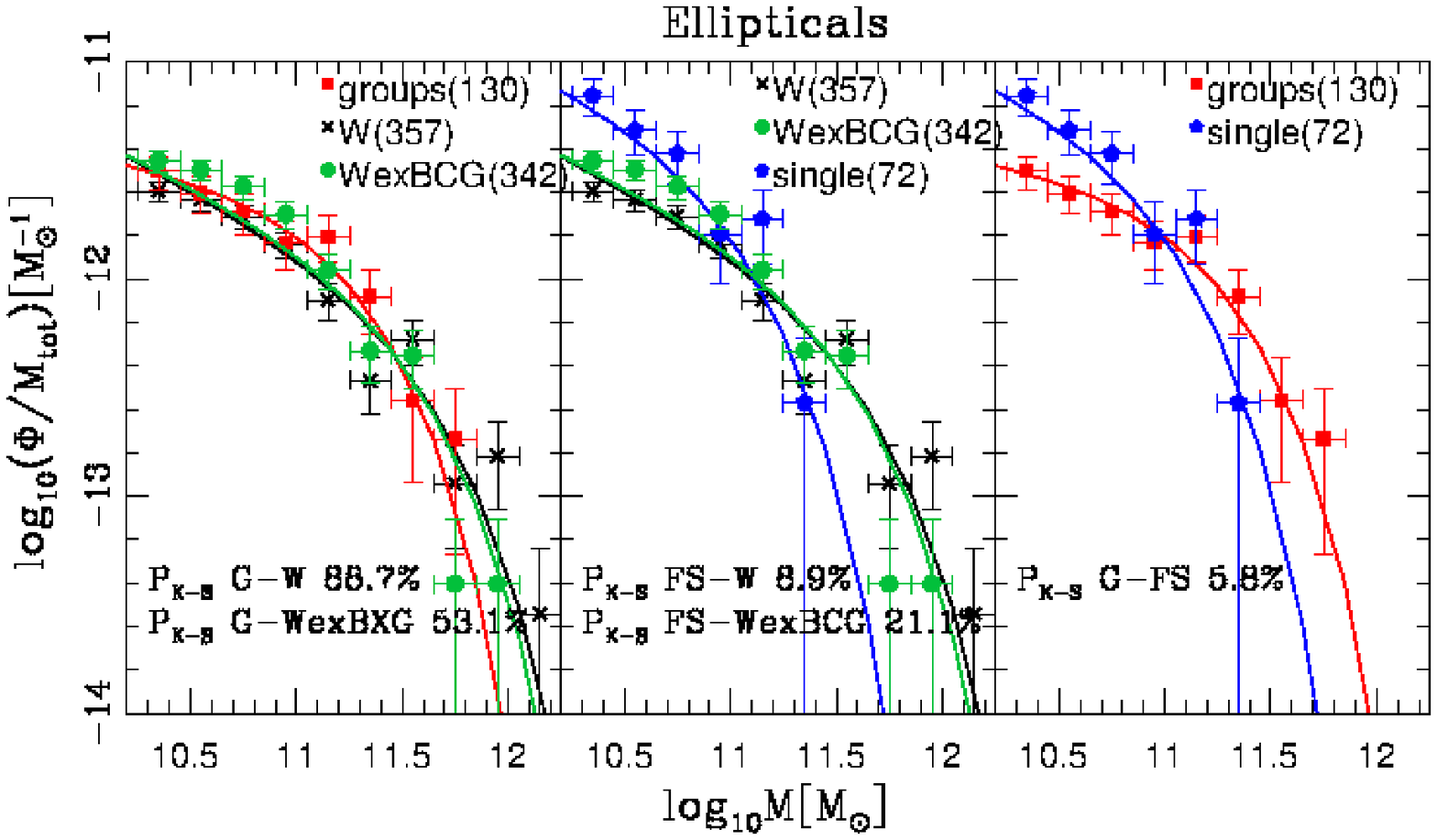}}
    {\includegraphics[scale=0.52]{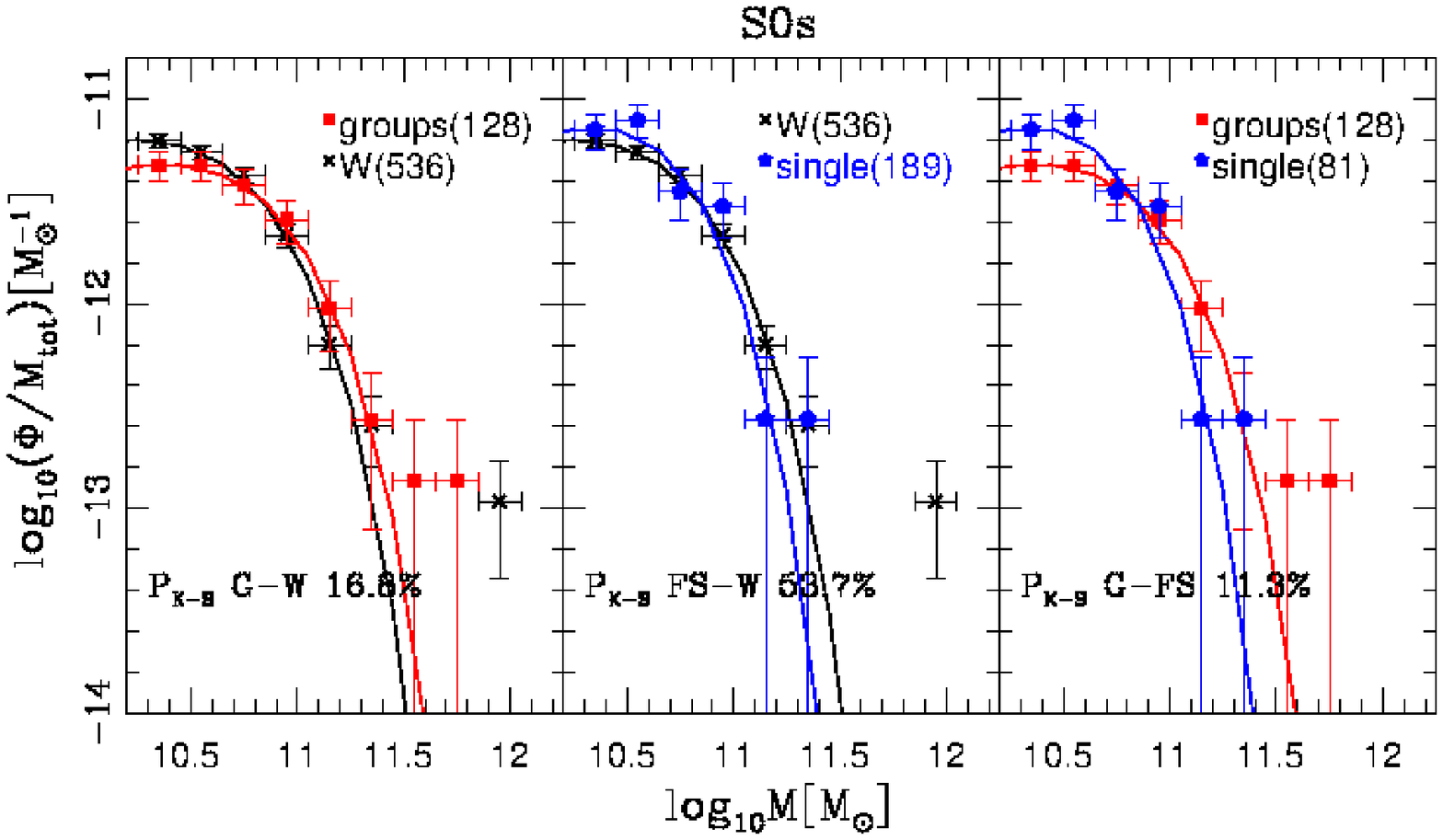}}
    {\includegraphics[scale=0.52]{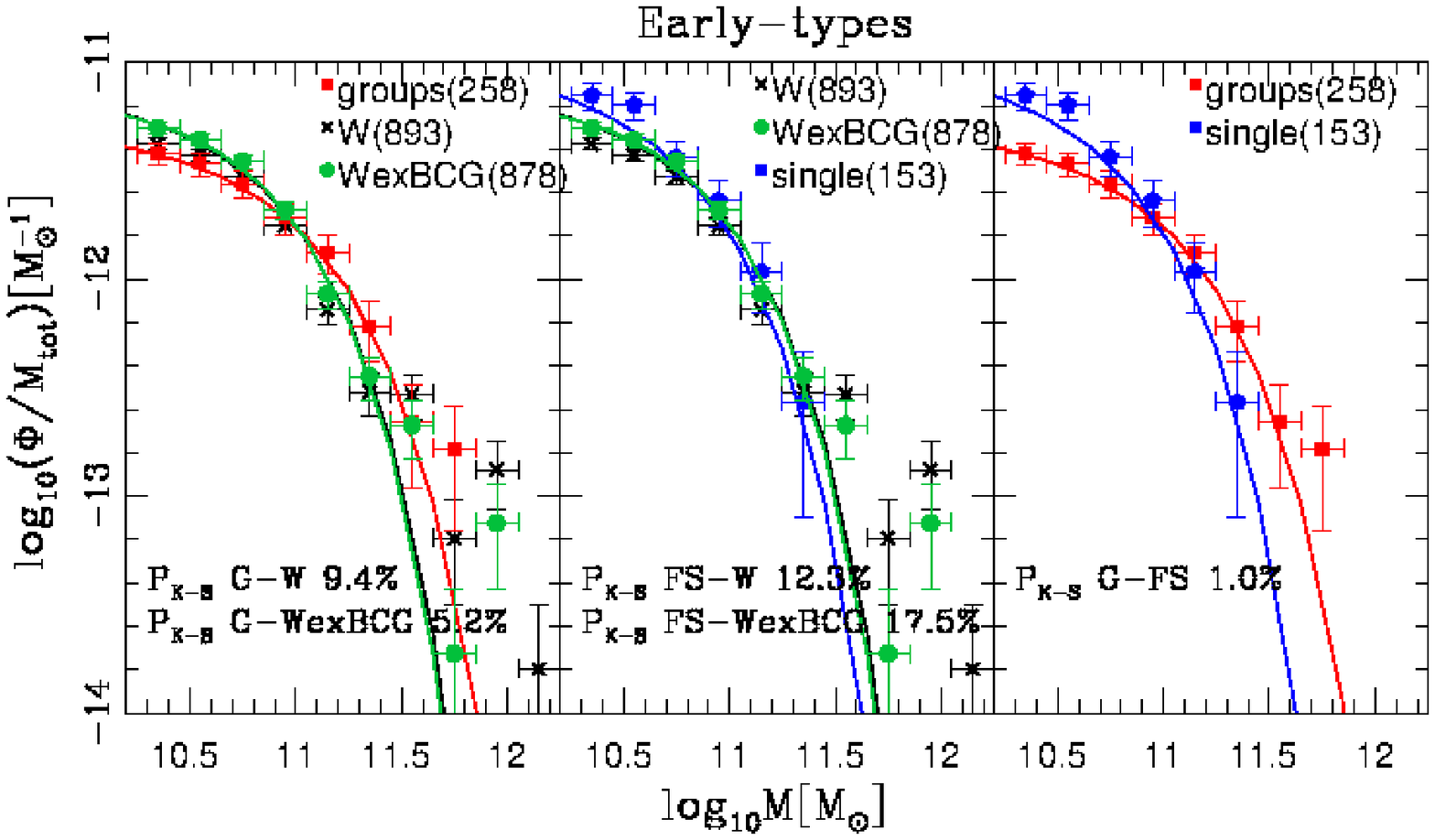}}
    {\includegraphics[scale=0.52]{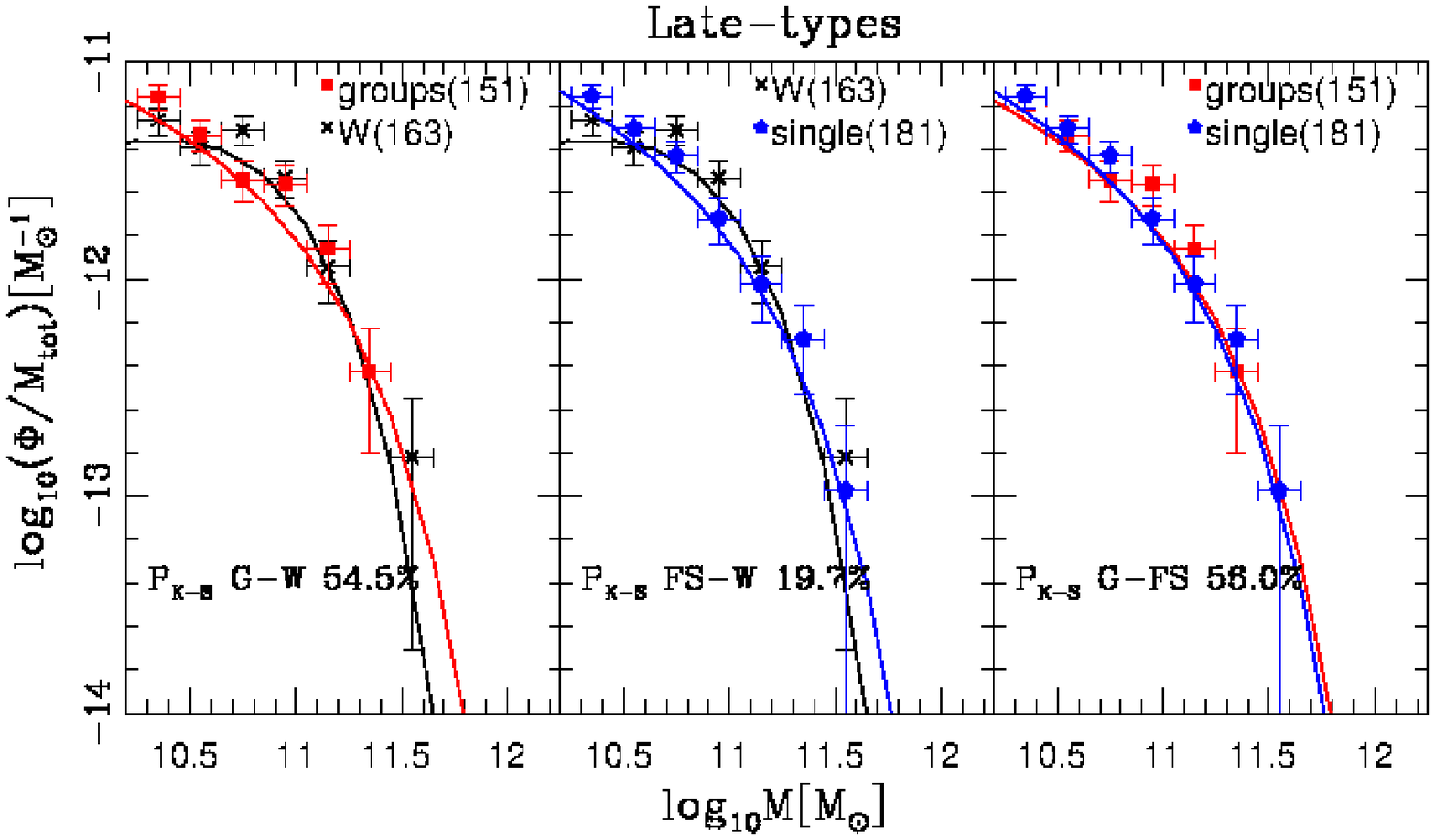}}
\caption{Mass distribution for different pairs of
environments for ellipticals, S0s, early-types and late-types. Errors
are defined as poissonian errors in the \textit{y} direction and are equal to
the bin size in the \textit{x} direction. Mass distributions are
normalized using the total integrated stellar mass, above the mass
completeness limit. Numbers in the brackets are the number of
galaxies above the mass limit and are weighted for WINGS.
}
\label{F4}
\end{figure*}

As for the total galaxy MFs, now we investigate the
variation of the MF of elliptical, S0, early and late type
galaxies in clusters, groups, binaries and singles.  Fig.~\ref{F4}
shows the distributions of each morphological class. Subdiving
our samples in both morphological type and detailed environment, the
statistics get worse, and in most cases the errors on the Schechter
parameters become too large to draw robust conclusions. In particular,
the binary sample is always too poor to be compared with the
others, and is not included in the following analysis.
Schechter parameters are anyway listed for completeness for
all environments in Table~\ref{sc_par}.

Comparing clusters and groups, which are the two environments with the
best statistics, we find small differences in the MF of
ellipticals, as seen in Fig.~\ref{F4}, from the K-S and the
Schechter fits.  Ellipticals in the single sample, instead, show a
steeper MF than those in groups and clusters, as seen in
the plot and, marginally, found by the K-S for groups.

The mass distribution of S0 galaxies in clusters and groups is
indistinguishable on the basis of the K-S test and of the Schechter
parameters, although the inspection of Fig.\ref{F4} shows a possible
steepening at low masses in clusters.
The S0 MFs of single
galaxies is too noisy to draw secure conclusions, but the plot
is suggestive of a steep fall-off at high masses.

For the early-type MF
small differences start
to be appreciable between clusters and groups especially
when excluding cluster BCGs: both the KS and the Schechter
$\alpha$ show differences at the 1$\sigma$ level.
Moreover, the differences in the MF of singles and groups
are now statistically significant and differences singles-clusters
are clearly visible in the plot.

Finally, we consider the mass distributions of late-type galaxies.
The shape of their MF is very similar in groups and singles,
while in clusters there is a flattening at masses
$\rm log_{10}M_{\star}/M_{\odot}< 10.65$, corresponding to a much
higher Schechter $\alpha$ value. This may correspond to the
steepening in the MF of cluster S0s at these masses,
if preferentially low-mass late-types are transformed in S0s by the
cluster environment.


To summarize, the MFs of ellipticals and S0 galaxies show
small differences between clusters and groups, while their
distributions appear much steeper in the single galaxy sample.
Therefore, isolated ellipticals have on average lower
masses than cluster and group ellipticals.

The mass distribution of late-type galaxies is similar in all
environments, except for a deficit of low-mass late-types in
clusters. These environmental variations are consistent with those
found in the previous section between general field and clusters,
where we observed an excess of massive ellipticals in clusters
compared to the general field, obviously driven by the steep high-mass
fall-off of the elliptical MF in single galaxies, and the deficit of
low-mass late-types in clusters.


\section{Discussion}

Our most important result is the intriguingly weak environmental
dependence of the galaxy stellar MF.  Above $\rm log_{10}
M_{\star}/M_{\odot} = 10.25$, the MF in the general field,
and in groups -- which are the dominant component of the general field
-- is similar to that in clusters.

It is important to emphasize that our sample consists of galaxies with
masses at least half of our Milky Way, and that stronger variations of
the MF with global environment can exist at lower galaxy
masses than those considered in this study.

Our results disagree with the conclusions of \citet{Bal11}, while
agree with \citet{vdL}.

\citet{Bal11} found a much higher $M^*$ and a much lower
$\alpha$ in clusters than in the general field in a sample that uses
2MASS photometry and Las Campanas Redshift Survey spectroscopy, in
which clusters are identified as structures with $\sigma > 400 \rm \,
km \, s^{-1}$ using a friends-of-friends algorithm.  It is difficult
to assess whether this difference is driven by the BCG cluster
galaxies, also because the MFs are not shown in Balogh et
al. What we find is that their $log M^*/M_{\odot} \sim 11.35$ value
(in our units and IMF) and $\alpha \sim -1.7$ (from their Fig.~12, for
an unknown galaxy mass limit) seem incompatible with our WINGS mass
function.

Instead, \citet{vdL}, based on SDSS data, studied
$\sim 500$ clusters and found no evidence for mass segregation in
clusters: excluding BCGs, using median masses and cumulative radial
distributions in mass bins, they found no evidence for a mass
dependence on clustercentric distance, out to $>10$ cluster virial
radii into the field.


In this paper we have analyzed the global environment. It is well
known that the galaxy stellar MF depends on local
environment, i.e.  on local galaxy density \citep{BA2,Bam},
and this is true also for the PM2GC sample used
in this paper, measuring local density to the 5th nearest neighbour, 
as discussed in \citet{VU2}. The fact that,
within the same sample, local density effects are much more easily
detectable than global environment effects suggests that the mass
distribution of galaxies is more strongly depending on local scale
processes than global ones such as halo mass (see discussion in
\citealt{VU2}).  A detailed comparison with the MFs
expected from galaxy simulations may help clarify the origin of these
differences, and is currently underway (Vulcani et al. in prep.).

%

\section{Summary}

We have analyzed the low-z stellar MF of galaxies with
masses $\rm log_{10} M_{\star}/M_{\odot} \geq 10.25$ in
different global environments, from clusters, to groups, binary and
single galaxies, and the general field.

The main result of our work is the overall striking independence of the shape of
the MF on global environment. Contrary perhaps to most expectations,
the MF in the general field is indistinguishable from that in
clusters. The cluster and the group MFs are also very similar, with
only subtle differences allowed at best.

The only environment where the MF differs significantly is the sample
of single galaxies, representing the lowest mass haloes containing the
most isolated galaxies and comprising about one third of the general
field population. The single galaxy MF is steeper, proportionally
richer in lower-mass galaxies, than other environments.

What varies with global environment is the maximum mass reached by
galaxies: the upper MF cut-off varies from 1.6-3.5 $\times 10^{11}
M_{\odot}$ in binaries and single systems, to 5.6 $\times 10^{11}
M_{\odot}$ in groups and $10^{12}$ in clusters, even excluding the
cluster BCGs. In line with theoretical expectactions,
this indicates that the most massive galaxies are only formed in the
most massive environments/haloes.

We stress that a stronger dependence of the MF on environment may of
course exist at lower galaxy masses than those considered in this
study, and that our sample includes galaxies down to masses about half
of our Milky Way.

Our results resemble those at higher redshifts of \citet{VU3},
that found a similar MF in clusters, groups and general
field at $z=0.4-0.8$, and a surprisingly similar evolution of the MF
in clusters and the field between $z \sim 0.8$ and today.
Their study could not discriminate single galaxies.
In fact, our results are also in agreement
with those at intermediate
redshifts of \citet{KOV10}, that found a difference in the mass
function of group and isolated galaxies (their Fig.~7).

In the second part of our paper we have presented the MF of
different morphological types (ellipticals, S0s and late-types) in
different environments. We have shown how the MF changes
from one type to another in each environment, and that the mass
function of a given morphological type may vary with
environment. The strongest environmental dependence is for
the (massive) ellipticals, while only very small differences
are observed for spirals.
These findings imply that both galaxy mass and
environment must play some role in establishing the distribution of
morphological types we observe in the local Universe.

\section*{Acknowledgments}
We thank Joe Liske, Simon Driver and the whole MGC team
for making easily accessible a great dataset. We are grateful to the
rest of the WINGS team for help and useful discussions.

\end{document}